\documentstyle[12pt]{article}
\def\by#1#2{{\displaystyle {#1}\over \displaystyle {#2}}}
\def\bynd#1#2{{{#1}\over {#2}}}
\def\d{{\rm d}}

\def\gev2{\hbox{GeV}^2}
\def\<{\langle}
\def\>{\rangle}
\begin{document}

\begin{flushright}
June 13th, 1996 \\
DO-TH-96-9
\end{flushright}

\begin{center}
{\Large \bf
Cancellation of infrared divergences at finite temperature} \\ [1cm]
D. Indumathi, \\ [0.3cm]
{\it Dept.of Theoretical Physics, University of Dortmund, D 44221,
Germany.} \\[1cm]
\end{center}

\begin{abstract}

We consider a typical hard scattering process in a heat bath of photons
and electrons at temperature, $T$, in finite temperature QED. We show
that, when the hard scattering scale is much larger than the
temperature, the infrared pieces of both the real and virtual parts of
the cross section factorise; these can be exponentiated and cancel
between each other to all orders in perturbation theory. We use the
technique of Grammer and Yennie to give a prescription for the
extraction of the infrared divergent parts, and for the form of the
finite remainder. Symmetry arguments are used to show the finiteness of
new terms arising in the $T \neq 0$ part of the computation. 

\end{abstract}

\vspace{1cm}

\section{Introduction}

The study of gauge field theories at finite temperature has been of
interest for several years \cite{bkgnd}, both for the physics of the
early universe, as well as for the understanding of Quark Gluon Plasma
(QGP) \cite{qgp}. In particular, finite temperature Quantum
Chromodynamics (QCD) is relevant in order to examine the possibility of,
and evidence for the formation of QGP in relativistic heavy ion
collisions. A lot of results also exist with Quantum Electrodynamics
(QED), rather than QCD, as the theory at hand, due to the simpler nature
of the abelian theory. The infra-red (IR) finiteness of such theories is
not generally established; however, the manner of extraction of an IR
finite result can vitally affect many finite quantities, for example,
correlations between final state momenta in a hard scattering process at
finite temperature \cite{Gupta}. 
Only a few calculations in finite temperature QCD or in non-gauge
theories have actually demonstrated the IR finiteness of the process
under study, to the order calculated \cite{misc}. Recently, Weldon
\cite{Weldon} showed the IR finiteness of the simpler problem of hard
scattering in thermal QED to all orders, within the eikonal
approximation. 

This suppression of IR divergent terms in a hard scattering of a
classical charge with, say, an off-shell photon of momentum $q$, in a
heat bath of thermal photons at temperature, $T$, $-q^2 \gg T^2$, takes
place as follows:

A hard scattering occurring in a heat bath is accompanied by soft photon
emission and absorption. It is therefore proper, when considering higher
order corrections to this process, to include not only the virtual
corrections, but the real (unobserved) soft photon contributions as
well. Weldon
showed that, in the eikonal approximation, the 1--loop virtual correction
to such a hard scattering cross section factorises as the product of the
leading order (hard) scattering cross section times an IR--divergent
factor, which contains the square of the semi-classical current:
$$
{\widetilde{J}}_\mu(k) = \left(\by{p'_\mu}{p' \cdot k} -
\by{p_\mu}{p \cdot k} \right)~,
\eqno(1)
$$
along with some thermal factors. Here $k$ is the loop momentum and $p$
($p'$) is the initial (final) momentum of the electron that
participates in the hard process, and we have suppressed a factor of
$ie$ for later convenience. The $n$-loop result can be expressed in
terms of the 1--loop result; the all-order virtual contribution is thus
the exponential of the 1--loop result. 

The contribution from the corresponding order {\it real} photon emission
and absorption (with respect to the heat bath) diagrams, on performing
the photon polarisation sum, contains the same ${\widetilde{J}}^2$
factor; furthermore, the total cross section, to this order,
which is the sum of these two real and virtual parts, turns out to be IR
finite. By expressing the $n$-photon process as the product of $n$
similar factors, each corresponding to 1-photon emission/absorption,
Weldon was able to express the all-order result as an exponential
of the leading order correction times the hard scattering cross section;
such a factorisation occurred because of the use of the eikonal
approximation. Weldon thus
demonstrated the IR finiteness of such a thermal hard scattering cross
section by showing the cancellation between the real and virtual
contributions in the exponent.

In this paper, we re-examine the infrared behaviour of the same cross
section {\it without making the eikonal approximation} and show that the
exact cross section is indeed also IR finite to all orders. Hence the
use of the eikonal approximation is valid in calculations of this type. 
We have obtained such a proof of IR finiteness using the technique of
Grammer and Yennie (GY) \cite{GY} to identify and separate the IR
divergent pieces. This involves separating the contribution of the
soft photon propagator into the sum of two parts---called $K$ and $G$
photon contributions respectively---in the matrix element corresponding
to the virtual soft photon corrections, and separating the polarisation
sum in the cross section for the real soft photons into two parts---the
$\widetilde{K}$ and $\widetilde{G}$ contributions respectively. The $K$
($\widetilde{K}$) parts of the virtual (real) contributions
contain the IR divergence; the $G$ ($\widetilde{G}$) photons give
finite contributions.

The IR divergent part of the virtual contribution arising from the $K$
photons can be expressed in terms of the quantity,
$$
{J}_\mu(k) = \left(\by{(2 p'- k)_\mu}{(k^2 - 2 p' \cdot k)} -
\by{(2 p- k)_\mu}{(k^2 - 2 p \cdot k)} \right)~,
\eqno(2)
$$
just as was obtained by GY for the corresponding $T = 0$ case. However,
now there are extra thermal factors involved as well, so that the
projected-out IR divergent piece differs from the $T=0$ result of GY.

The IR-divergent virtual contribution, to all orders, that multiplies
the matrix element corresponding to the hard scattering cross section
then turns out to be expressible as the exponential of the 1--loop
result, viz., as $\vert \exp [{B}] \vert^2$, where 
$$
{B} = \by{i e^2}{2(2\pi)^4} \int \d^4 k {J}^2 \left[ \by{1}{k^2 +
i \epsilon} - 2 \pi i \delta(k^2) N (\vert {\bf {k}} \vert) \right]~,
\eqno(3)
$$
and the Bose distribution function, $N$, for thermal photons, is given by,
$$
N(\vert {\bf {k}} \vert) = \left(\exp (\vert {\bf {k}} \vert /T)
                            - 1 \right)^{-1}~.
\eqno(4)
$$
The IR divergent part of the real photon cross section (the
${\widetilde{K}}$ contribution to all orders) can also be
factorised and expressed as the exponential of the corresponding
1--photon result, viz., as $\exp[\hat{B}]$, where
$$
\hat{B}(x) = - \by{e^2}{(2\pi)^3} \int \d^4 k \delta(k^2)
{\widetilde{J}}^2 \left[ \theta(k_0) + N(\vert {\bf {k}} \vert) \right]
\exp[ i k \cdot x]~.
\eqno(5) 
$$
Both $B$ and $\hat{B}(x)$ contain logarithmic as well as linearly
diverging terms in the IR limit. This is because both the currents, $J$,
and $\widetilde{J}$, and the Bose distribution function, $N$, have a
behaviour $\sim 1/k$ for small $k$. Then the IR finiteness of the total
cross section is demonstrated by showing that the IR divergent parts of
$(B + B^*)$ and $\hat{B}$ are equal and opposite in sign. There are two
steps in the proof: (1) to factorise out the IR divergent piece in terms
of either $J^2$ (for the virtual diagrams) or ${\widetilde{J}}^2$ (for
the real part) such that they can be expressed as the exponential of
the corresponding 1--loop result, and (2) to show the cancellation of
these IR divergent exponentials between the real and virtual parts. 
It is a remarkable result that, while the logarithmic divergences (at $T
= 0$) and the linear divergences (at $T \neq 0$) factorise and can be
exponentiated and cancelled between the real and virtual contributions
to the cross section, the sub-leading terms (at $T \neq 0$) which
initially seem logarithmically divergent are also rendered finite due to
the symmetric nature of the $T \neq 0$ contribution and so do not spoil
the Bloch-Nordsieck theorem \cite{BN} at finite temperature. 

In the next section, we set up the notation and define the particular
hard scattering problem we wish to study. In Section 3, we use the
technique of Grammer and Yennie to show the factorisation and subsequent
exponentiation of the IR divergent pieces for virtual photonic
corrections to the hard scattering problem. We show that the IR
divergences from virtual photons arise from the $K$ photons whereas the
$G$ photons contribute IR finite pieces. We repeat the
analysis for real photon emission and absorption; viz., we show that the
$\widetilde{G}$ photon contribution is IR finite and the IR divergence
is contained in the $\widetilde{K}$ photon contributions, and can be
exponentiated. We then sum the real and virtual contributions
to arbitrary order; we show that the IR divergent terms cancel between
the real and virtual contributions, so that the resulting cross section
is IR divergence-free. In Section 4, we put back the electrons, which we
had so far neglected, into the heat bath (and hence include thermal electron
loops as well in the calculation) and show that electrons cannot be a
source of IR divergence. We conclude the paper in Section 5.
Finally, Appendices A, B, and C contain some details of various parts of
the calulations for virtual photon insertion, real photon insertion, and
thermal fermions respectively. 

The present work follows closely the proof technique of IR-finiteness in
$T = 0$ QED by Grammer and Yennie; hence the original technique is
introduced and explained as necessary in each section. For the sake of
completness, a lot of the results that were already obtained by them at
$T = 0$ are re-established here. This makes the extensions to, and the
differences from the $T \neq 0$ case easier to understand, as well as
provides for continuity and cohesion in this present work. 

We begin by defining the problem of interest. 

\section{The Process: Definitions and Notations}

We discuss the hard scattering of an electron with a photon in a fixed
temperature plasma of electrons and photons at temperature $T$.
For specificity, we
assume the hard process to be, 
$$
\gamma^* (q) \; e(p) \rightarrow e (p')~,
$$
that is, the scattering of an electron off an off-shell photon, with
$-q^2 \gg T^2$. In other words, all thermal fluctuations are soft so
that there are no hard thermal loops. The case
of $e^+ e^-$ annihilation into an off-shell photon could equally well be
described this way. The collision causes the charged particle to radiate.
Due to the presence of the heat bath, both absorption and emission of
photons takes place. 

We consider the generating functional for $T \neq 0$ QED in a real-time
formulation \cite{realtime}: 
$$
Z(j,T) = \int{\cal D} [A, \psi, \overline{\psi}]
\exp\left[i\int \d^4x\left( -{1\over 4}F_{\mu\nu}F^{\mu\nu}+
\overline{\psi} (i \not{\!\partial} - m) \psi +
j_\mu A^\mu\right)\right]~,
$$
where $j_\mu$ is usual electromagnetic current and the integration in
the complex time plane is over a contour that includes the temperature.
It is so chosen to obtain correct thermal
averages of the $S$-matrix elements \cite{Rivers}. The periodicity
condition on each of the fields, 
$$
\phi (t_0) = \phi (t_0 - i \beta); ~~ \beta = 1/T~,
$$
for any real $t_0$, results in the well-known field doubling, with
propagators assuming a $2 \times 2$ matrix form. The (11) component of
the propagators correspond to the usual free-field result. The type-1
fields are physical fields, while the type-2 fields are ghosts and
cannot appear on external legs. The off-diagonal components of the
propagator allow transmutations of one type into another. The photon
propagator corresponding to a momentum $k$ is then given by,
$$
\begin{array}{rcl}
i {\cal D}_{ab}^{\mu\nu} (k) & = & - g_{\mu\nu} i D_{ab} (k)~, \\
i D_{ab} (k) & = & \left(
\begin{array}{cc}
\Delta & 0 \\
0 & \Delta^* \\
\end{array}  \right) +
2 \pi \delta(k^2) N( \vert {\bf{k}} \vert )\,
\left(
\begin{array}{cc}
1 & {\rm e}^{\vert {\bf{k}} \vert /(2T)} \\
{\rm e}^{\vert {\bf{k}} \vert /(2T)} & 1 \\
\end{array} \right)
\end{array}
\eqno(6)
$$
where $\Delta = i/(k^2 + i \epsilon)$, 
while the electron propagator for an electron of momentum $p$ and in
zero chemical potential is given by,
$$
\begin{array}{rcl}
i S_{ab} (p, m) & = & \left(
\begin{array}{cc}
S & 0 \\
0 & S^* \\
\end{array} \right) + 
2 \pi \delta(p^2 - m^2) N_f( \vert {\bf{p}} \vert ) \,
\left( \begin{array}{cc}
1 & \epsilon(p_0) {\rm e}^{\vert {\bf{p}} \vert /(2T)} \\
- \epsilon(p_0) {\rm e}^{\vert {\bf{p}} \vert /(2T)}  & 1 \\
\end{array} \right)~,
\end{array}
\eqno(7)
$$
where $S = 1/(\not{\!p} - m + i \epsilon)$. The first term in each case
is the $T = 0$ piece and the second, the $T \neq 0$ part. 

Vertices only connect fields of the same type---all fields at the
vertex, $\mu$, are of type $\mu$. The corresponding vertex factor is
$(-i e)(-1)^{\mu + 1}$, so that the ghost field vertex differs by a sign
from the physical field vertex. This completes the rules required to
draw and compute the various Feynman graphs contributing to the hard
scattering process.

Note that the external hard photon is not thermalised with respect to
the heat bath and so the hard scattering vertex corresponds to a type-1
vertex alone; the vertices adjacent to the external (incoming and
outgoing) electron legs can also be of type 1-alone, since only physical
fields appear on external legs of Feynman diagrams.

Fermions in the thermal bath are distributed according to the
Fermi-Dirac distribution, 
$$
N_f(\vert {\bf{k}} \vert) = \left(\exp (\vert {\bf{k}} \vert /T)
                            + 1 \right)^{-1} \stackrel{k \to
			    0}{\longrightarrow} \by{1}{2}~,
\eqno(8)
$$
which tends to a constant in the IR limit, unlike the Bose distribution,
which adds an inverse power of $k$. Hence thermal photons, and not
fermions, determine the IR behaviour of the finite temperature cross
section. The presence of electrons in the heat bath then provides an
inessential complication in the problem of the IR behaviour of such
cross sections. We therefore begin by ignoring the thermal nature of
electrons, and consider only their $T = 0$ parts. It is then sufficient
to discuss only the (11) component of the photon propagators and
disregard field doubling. We shall do this in the next section. After
demonstrating the IR finiteness of the hard scattering cross section
in this limit, we re-instate the electrons in the heat bath in Section
4, and show that our results still hold. For the purposes of calculation
in the next section, then, the (11) photon propagator is given below:
$$
i D_{11}^{\mu\nu}(k) = - i g_{\mu\nu} \left[ \by{1}{k^2 + i \epsilon}
- 2 \pi i \delta(k^2) N(\vert {\bf{k}} \vert) \right]~,
\eqno(9) 
$$
or, in a Lorentz covariant manner, by replacing $\vert {\bf{k}} \vert \to
k \cdot u$, where the Lorentz vector, $u$, specifies the 4--velocity of
the heat bath with respect to the observer \cite{Gupta,Weldonold}. We
see that finite temperature effects are felt only on mass shell.

Apart from modifications to the propagator, as shown above, the finite
temperature affects the phase space for real photons as well.
This is because photons can be emitted into as well as absorbed from the
heat bath. In fact, without including absorption diagrams, the IR
divergences will not cancel \cite{Gupta,Weldon}. This can be taken into
account by writing the phase space element corresponding to the $i$-th
photon field of type-1 as \cite{Gupta}
$$
\d\phi_{k_i}\;=\;{d^4k_i\over(2\pi)^4} 2\pi\delta\left(k_i^2\right)
 \left[\Theta\left(k_i^0\right)+N(\vert {\bf{k}} \vert)\right]~.
\eqno(10) 
$$
Emission of particles corresponds to $k_i^0 > 0$ and absorption to
$k_i^0 < 0$. This incorporates the correct statistical factors of $(1 +
N)$ for emission and $N$ for absorption of photons. The appearance of
the thermal factor of $N$ in the phase space worsens the IR behaviour in
comparison with the corresponding $T = 0$ problem, since $N \sim 1/k$. 

Hence we see that the finite temperature contribution coming from real
as well as virtual photons have each a $T = 0$ and a $T \neq 0$ part. 
Note that an ultraviolet (UV) cut off on $k$ is required in the $T=0$ case
\cite{Gupta}. However, the presence of the Bose distribution function
obviates the need for such a cut-off in the $T \neq 0$ case. 

\section{The Cross Section}

We are now ready to compute the soft corrections to the hard scattering
process coming from both virtual diagrams (electron self energy and
vertex corrections) as well as real photon emissions and absorptions.

We use the approach of Grammer and Yennie (GY) \cite{GY} who developed
a method to separate the IR divergent and
IR finite expressions in cross sections in QED at zero temperature. The
crux of their technique is the separation of the photon propagator for
virtual photons (or the photon polarisation sum for real photons) into a
sum of two modified polarisation sums called $K$ and $G$ type sums. The
$G$ sum contains the IR finite part while the $K$ sum contains the IR
divergent part and will conveniently factor out of the cross section.
We will show that such a factorisation occurs for $T \neq 0$ as well. 
These $K$ terms depend only on the momenta of the external charged
particles and can be easily exponentiated. On summing the real and
virtual corrections, the result in IR finite. We begin with the virtual
photon contribution. 

\subsection{The virtual photons}
The rearrangement is accomplished, for virtual photons, by replacing, in
the photon propagator, eq. (9), the factor $g_{\mu\nu}$ by
$$
\begin{array}{rcl}
g_{\mu\nu} & \to & (g_{\mu\nu} - b k_\mu k_\nu) + b k_\mu k_\nu \\
& \equiv & G_{\mu\nu} + K_{\mu\nu}~.
\end{array}
\eqno(11)
$$
$k$ is the photon loop momentum, as usual, and $b$ is a $k$ dependent
factor that depends on how the photon is introduced into the
graph\footnote{Note a slight modification in $b$ with respect to the
original definition of GY for $p_f \neq p_i$; this is needed in order
to remove logarithmic sub-divergences that occur at $T \neq 0$.}:
$$
b(p_f, p_i) = \by{1}{2} \left[\by{(2p_f - k) \cdot (2p_i - k)}
       {(k^2 - 2 p_f \cdot k)(k^2 - 2 p_i \cdot k)}+
        (k \leftrightarrow -k) \right]~.
\eqno(12)
$$
For large $k$, this corresponds to choosing the Landau gauge. 
Here $p_i$ ($p_f$) is equal to $p$ or $p'$ according to whether the
initial (final) point of the inserted photon is in the $p$ or $p'$ leg
(i.e., is attached to the initial or final electron line). 

Hence every virtual photon contributes either as a $G$ or $K$ type
photon. Notice that the factorisation is independent of the temperature;
in fact, it also holds for all components of the photon propagator
matrix in the complete interacting theory: we will make use of this
fact in the next section. 

Finally, to account for the external line renormalisations, GY use the
following rule: omit both the wave function renormalisations as well as
drop self-energy corrections in the outgoing electron line. We adopt
this procedure as well.

\subsubsection{$K$ photon insertion}
\begin{figure}[ht]

\vskip 7truecm

\includegraphics{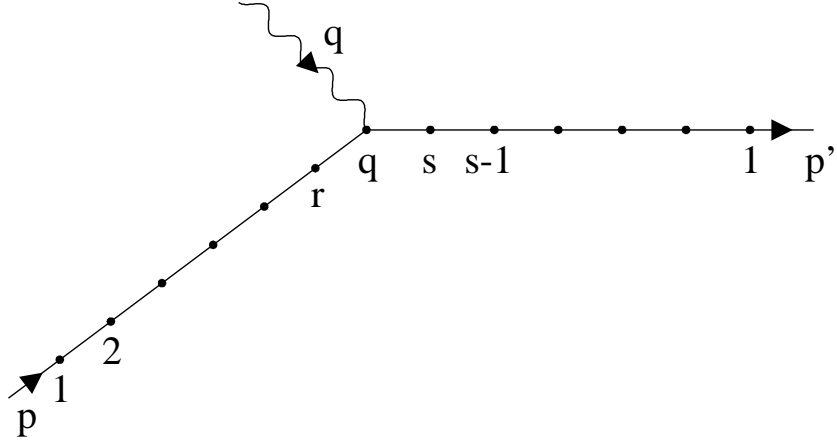}
\caption[dummy]{\small A graph with $n$ photon vertices, labelled from
$1$ to $r$ on the initial or $p$ electron leg, and from $s$ to $1$ on the
final or $p'$ leg, $r+s=n$. The attached photons can be either real or
virtual. This is the starting point for the analysis
of an additional $(n+1)$-th $K$ or $G$ type photon.}
\label{fig1}

\end{figure}

Let us look at the effect of inserting a $K$ photon of momentum
$k_{n+1}$ into a graph containing $n$ photon vertices. Since this
corresponds to the replacement, eq. (11), 
$$
g^{\mu\nu} \to b_{k_{n+1}} k_{n+1}^\mu k_{n+1}^\nu~,
$$
where $b$ is defined in eq. (12), such an insertion has the factor
$\not{\!k}_{n+1}$ at the insertion vertices. It is most convenient to
work with completely symmetric graphs, that is, we consider not only an
$n$-photon graph, but the set of those graphs obtained by all possible
permutations of the $n$ indices, and compensate for the overcounting by
dividing the result by $n$! in the end. A typical graph is shown in
fig.\ \ref{fig1}. When the additional $K$ photon is inserted into this
graph, there is a factor $\not\! k_{n+1}$ at each point of insertion,
sandwiched between two electron propagators with momenta $p_r$ and
$p_r - k_{n+1}$. The relevant part of the matrix element reads,
$$
\by{1}{\not{\!p}_r - \not{\!k}_{n+1} - m} \not{\!k_{n+1}}
\by{1}{\not{\!p}_r - m}
$$
and this can be simplified to a difference of two terms:
$$
\by{1}{\not{\!p}_r - \not{\!k}_{n+1} - m} \not{\!k_{n+1}}
\by{1}{\not{\!p}_r - m} = 
\by{1}{\not{\!p}_r - m}  \not{\!k_{n+1}}
\by{1}{\not{\!p}_r - \not{\!k}_{n+1} - m} = 
\by{1}{\not{\!p}_r - \not{\!k}_{n+1} - m} - 
\by{1}{\not{\!p}_r - m} ~,
\eqno(13)
$$
using Feynman's identity. The $(n+1)$-th photon must be inserted at all
possible points; this results in a pairwise cancellation of terms, due
to the above identity, leaving only the contributions from the last and
first insertions, which fail to cancel (only all such graphs must be
considered which have the same overall factor, $b(p_f, p_i)$). This
leads to a great simplification in the computation of these
contributions. We demonstrate this cancellation explicitly in Appendix A. 

\vspace{0.3cm}

Then, following closely the arguments of GY, we see that the addition of
a virtual $K$ photon in all possible ways results simply in an overall
factor multiplying the matrix element of the original $n$-photon graph
(for details, see Appendix A);
this factor (writing $k_{n+1} \equiv k$, for simplicity) is,
$$
\begin{array}{rcl}
 & \sim & \displaystyle \by{i e^2 }{2 (2\pi)^4} \int \d^4 k
 \left[b(p, p) + b(p', p') - 2 b(p', p)\right] \left[ D_{11} (k) 
 \right]~, \\
 & = & \displaystyle \by{i e^2 }{2 (2\pi)^4} \int \d^4 k
 \left[{J}^2\right] \left[\by{1}{k^2 + i \epsilon} - 2 \pi i 
 \delta(k^2) N (\vert {\bf{k}} \vert) \right]~, \\
 & \equiv & {B}~, 
\end{array}
\eqno(14)
$$
where ${J}_\mu$ is defined in eq. (2) and ${B}$ in eq. (3). Note that
$J$ reduces to $\widetilde{J}$ in the small $k$ limit. $B$ exhibits both
logarithmic and linear divergences in the IR, as stated earlier. We now
turn our attention to $G$-photon insertions in the symmetric $n$-photon
graph. 

\subsubsection{$G$ photon insertion}

The $G$ photon contribution corresponds to re-writing the
$g_{\mu\nu}$ term in the photon propagator (see eqs. (9), (11)) as
$$
g_{\mu\nu} \to g_{\mu\nu} - b_k (p_f, p_i) k_\mu k_\nu~.
$$
We wish to show that the $G$ photon contribution is IR finite. In the
original GY paper (for $T = 0$ QED), this was done by showing that the
IR-divergent parts of $g_{\mu\nu}$ and $b k_\mu k_\nu$ terms cancel,
leaving behind a finite remainder. The crux of the solution lay in
considering a typical contribution when all powers of $k$ were dropped
in the numerator: this is the most divergent term. The matrix element
for the $n$-photon graph with an
additional virtual photon insertion at vertices $\mu$ and $\nu$ is,
$$
{\cal M}^{n+1} \sim \overline{u}_{p'} \cdots (\not{\!p}_f+m) \gamma_\mu
(\not{\!p}_f+m) \cdots (\not{\!p}_i + m) \gamma_\nu (\not{\!p}_i + m)
\cdots u_p + \ldots ~,
\eqno(15)
$$
where the last set of ellipses indicate terms with factors of $k$ in
the numerator, and we have not shown the denominators. The presence of
the external spinors, as well as the fact that $p_f$ and $p_i$ satisfy
the mass-shell condition (they are one of $p$, $p'$) enables us to
express the leading term  as,
$$
\begin{array}{rcl}
{\cal M}^{n+1} & \sim & \overline{u}_{p'} \cdots (\not{\!p}_f+m)
2 p'_\mu \cdots 2 p_\nu (\not{\!p}_i + m) \cdots u_p~, \\
 & \propto & p'_\mu p_\nu~.
\end{array}
$$
Hence, the $G$ photon contribution can be expressed as
$$
\begin{array}{rcl}
{\cal M}^{n+1} & \sim & p_{f,\mu} p_{i,\nu} \left\{g_{\mu\nu} -
b_k (p_f, p_i) k_\mu k_\nu \right\}~, \\
 & = & 0 + {\cal O} (k)~,
\end{array}
\eqno(16)
$$
the vanishing of the leading divergence occurring due to the definition
of $b$ (see eq. (10)). Since the graph was originally logarithmically
divergent at $T = 0$, the surviving terms in eq. (15), denoted by the
ellipses, which have at
least one Power of $k$ in the Numerator (one PkN) are IR finite. For $T
\neq 0$, the graph was originally linearly divergent in the IR. The
cancellation of the leading divergence thus ensures that the linearly
diverging pieces of $g_{\mu\nu}$ and $b_k k_\mu k_\nu$ cancel;
however, there are left over (potentially) logarithmic subdivergences
coming from the $T \neq 0$ terms in eq. (15) even with respect to the
leading term. This is the major difference between the $T = 0$ and the
$T \neq 0$ calculations. These terms arise from the
${\cal O} (k) $ terms in eq. (16)
which are linear in $k$. We now make use of the
symmetry in $b$ with respect to $k \to -k$; and also recall that the
$T \neq 0$ terms occur only on the mass shell, i.e., when $k^2 = 0$.
The factor $b$ then reduces to,
$$
\begin{array}{rcl}
b^{T\neq 0}(p_f, p_i) & = & \by{1}{2} \left[\by{(4p_f\cdot p_i -
		2 (p_f + p_i) \cdot k)} {(2 p_f \cdot k)(2 p_i \cdot k)}+
	       (k \leftrightarrow -k) \right]~, \\
	    & = & \by{ p_f \cdot p_i}{ p_f \cdot k \,p_i \cdot k}~.
\end{array}
$$
Hence, we see that the terms linear in $k$ in the numerator vanish on
symmetrising with respect to $(k \to -k)$ for the $T \neq 0$ terms, so
that the combination,
$$
p'_\mu p_\nu \left\{g_{\mu\nu} - b_k^{T \neq 0} (p_f, p_i)
 k_\mu k_\nu \right\} =  0~,
\eqno(16a)
$$
exactly, with no left over numerator powers of $k$. Hence, the only
divergence coming from the $T \neq 0$ terms is a linear one, and this is
cancelled by the choice of $G_{\mu\nu}$. 

We summarise the above result by stating that, when we retain only
$(\not{\!p}' + m)$ or $(\not{\!p} + m)$ terms in the numerator of the
matrix element corresponding to a $G$ photon insertion, the $T = 0$
contribution is IR finite because the leading logarithimic divergence
cancels between the two terms in $G_{\mu\nu}$. Terms containing one
or more PkN's in $b$ are anyway IR finite. For the $T \neq 0$ case, the
leading linear divergence is cancelled between the two terms of
$G_{\alpha\beta}$. Since the finite temperature effects are felt only on
mass-shell, the definition of $b$ can be further simplified by setting
$k^2 = 0$; the symmetry of the result in $k$ then ensures that there are
no PkN's in $b$; hence there can be no logarithmic divergences coming
from such terms, so that these terms are IR finite as well. Hence the
expression explicitly shown in eq. (15) is IR finite with respect to
$G$ photons. We now have to consider only those parts of the matrix
element which contain some powers of $k$ in the numerator, i.e., the
last set of ellipses in eq. (15). 

At $T = 0$, any PkN will render the result IR
finite. Since the leading divergence is linear due to the presence of
the Bose distribution function, this is not true at $T \neq 0$. A single
PkN can give rise to logarithmic
subdivergences, while higher powers of $k$ in the numerator yield an IR
finite result. However, just as the symmetry in the problem removed the
subdivergence with respect to PkN's in $b$, we
will find that a similar symmetry consideration can be used to soften
the logarithmic divergence coming from PkN's in
the matrix element, so that the $G$ photon contribution is IR finite at
all temperatures. We shall demonstrate this below. 

Before we do this, we review some notation and definitions needed for
this purpose. 

\subsubsection{IR finiteness of $G$ photons at $T \neq 0$}

We focus on the (logarithmic) subdivergences coming from factors of $k$
in the numerator of the matrix element at $T \neq 0$. 
To determine the divergence of a graph, we use power counting. Consider
the $n$-photon graph in fig.\ \ref{fig1} to which we wish to add a
$G$ photon, and reduce it to its skeleton (i.e., remove all divergent
subgraphs) \cite{GY}. The
set of $m$ photon lines that make up the skeleton form the controlling
set of the IR divergence. The divergence arises only when every $k_i, i
\in [1, m]$, simultaneously goes to zero. (There may be
one or two momenta in this set that give a logarithmic
divergence when they vanish (even when the other momenta in the set are
non-vanishing) with respect to the $(\not{\!p}' + m)$ or
$(\not{\!p} + m)$ terms in the matrix element. However, we have already
considered these terms and shown them to be IR finite with respect to
the insertion of $G$ photons. These momenta are however IR finite with
respect to the terms containing PkN's in the
matrix element, which is the case of interest here). We now show that
the symmetry of the matrix element in $(k_i \to - k_i)$ removes the
potential IR divergence arising from terms having one power of $k_i$ in
the numerator. This symmetry arises because $k$ is a loop momentum
corresponding to a virtual photon insertion. We assume, for simplicity,
that all the $m$ photons are $G$ photons and so the surviving divergence
arises only from the $T \neq 0$ piece. Later we shall relax this
assumption and prove the same result for the general case. 

\begin{figure}[ht]

\vskip 7truecm

\includegraphics{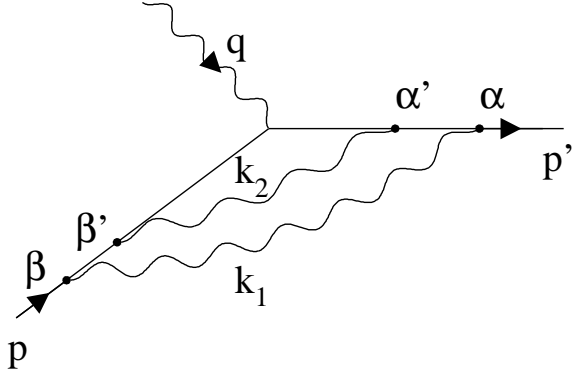}
\caption[dummy]{\small A second order ladder graph with controlling
momentum $k_1$.}
\label{fig2}
\end{figure}
\vspace{0.3cm}

We begin by demonstrating the cancellation of the logarithmic
subdivergence for specific lower order diagrams. For example, the second
order ladder graph in fig.\ \ref{fig2} can be expressed as,
$$
\begin{array}{rcl}
{\cal M}^{\rm Fig. 2} & \sim & \displaystyle \int \d^4 k_1 \d^4 k_2 
D(k_1) D(k_2) g_{\alpha\beta} g_{\alpha'\beta'} \times \\
& & \gamma_\alpha \by{\not{\!p}' - \not{\!k}_1 + m}{ a_1'} \gamma_{\alpha'}
\by{\not{\!p}' - \not{\!k}_1 - \not{\!k}_2 + m}{ a'_{12} - c_{12}} 
(- i e \!\! \not{\!\epsilon}_q) \by{\not{\!p} - \not{\!k}_1 - \not{\!k}_2 + m}{ a_{12}
- c_{12}} \gamma_{\beta'} \by{\not{\!p} - \not{\!k}_1 + m}{ a_1} 
\gamma_\beta~,
\end{array}
\eqno(17)
$$
where we have used $ a_i = 2 p \cdot k_i - k_i^2$, $a_{ij} = a_i + a_j$,
$c_{ij} = 2 k_i \cdot k_j$, 
and all primed quantities correspond to replacing $p \to p'$. There is
no IR divergence when $k_2 \to 0$ with $k_1$ fixed. Only when the
outermost photon momentum, $k_1$ (which is the controlling photon
momentum for such ladder graphs) vanishes, is there an IR divergence.
Let us therefore examine the behaviour of eq. (17) with respect to the
$k_1$ integration. The relevant part of the integral, which is the one
PkN term for the $T \neq 0$ part, is 
$$
I = \int \left[\d^4k_1 \d^4 k_2 \delta (k_1^2) N_1 \delta (k_2^2) N_2
\right] \by{k_1^\rho}{a'_1 (a'_{12} - c_{12}) (a_{12} -c_{12}) a_1}~,
\eqno(18a)
$$
where $N_i = N (\vert {\bf{k}} \vert)$. 
Since $k_i, i = 1, 2$ are loop integrals, $I$ should remain unchanged
under $(k_1 \to - k_1)$, and $(k_2 \to - k_2)$. The quantity in square
brackets is symmetric under this transformation; hence, we have,
$$
I \to I' = \int \left[ \cdots \right] \by{-k_1^\rho}{a'_1 (a'_{12} + c_{12})
(a_{12} + c_{12}) a_1}~.
$$
Since $I = I'$, we have $I = (I + I')/2$, i.e.,
$$
I = \displaystyle \int \left[\d^4k_1 \d^4 k_2 \delta (k_1^2) N_1
\delta (k_2^2) N_2
\right] \by{k_1^\rho}{a'_1 a_1} 
\by{1}{(a'_{12} - c_{12}) (a_{12} + c_{12})} \, \left\{ 
 \by{2 k_1 \cdot k_2}{(a_{12} - c_{12})} + 
 \by{2 k_1 \cdot k_2}{(a'_{12} + c_{12})} \right\}~,
\eqno(18b)
$$
which is linear in $k_1$. 
Hence we see that the symmetry under $k_i \to - k_i$ softens the
logarithmic divergence in $I$; the extra factor of $k_1$ coming from
$k_1 \cdot k_2$ in the numerator ensures that $I$ is IR finite when $k_1
\to 0$. Another way of understanding this result is as follows: when
$k_i \to 0$, $c_{ij}$ vanishes faster than $a_i$ or $a_{ij}$. If we drop
the $c_{ij}$ terms from the expression (18a), the integrand (with one
power of $k_1$ in the numerator) is odd under $k_i \to - k_i$
and so the integral actually vanishes. Since $c_{ij}$ is not exactly
zero, the integral does not vanish exactly, but vanishes as $c_{ij} \to
0$; hence the factor of $k_1 \cdot k_2 = c_{ij}$ in the numerator of
eq. (18b). The surviving part of the integral then turns out to be
IR finite;
the IR divergent piece not contributing since it is odd under this
transformation.

\begin{figure}[ht]

\vskip 7truecm

\includegraphics{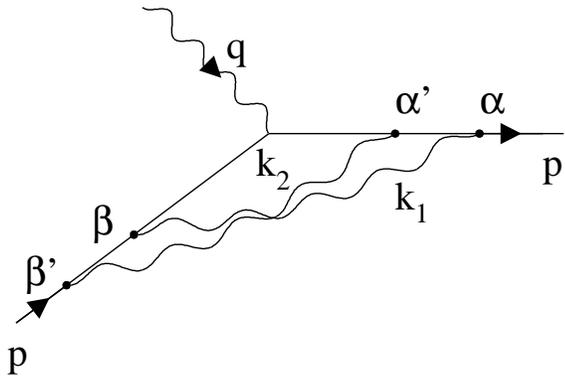}
\caption[dummy]{\small A second order ladder graph with controlling
momenta $k_1$ and $k_2$.}
\label{fig3}
\end{figure}
\vspace{0.3cm}

A similar analysis can be done for the cross graph of fig.\ ref{fig3}.
Here, both
$k_1$ and $k_2$ form the controlling set: the integral is divergent only
if both of them vanish simultaneusly. The leading contribution from the
$T = 0$ part is finite because of the choice of the combination of $G$;
any powers of $k_i, i = 1, 2$, in the numerator make the integral IR
finite. The leading (linearly divergent) contribution from the $T \neq
0$ piece is also finite due to the definition of $G$; we again have to
consider only the term with either one power of $k_1$ or $k_2$ in the
numerator of the $T \neq 0$ piece. There is again the factor,
$$
\by{1}{(a'_{12} - c_{12}) (a_{12} - c_{12})}~,
$$
in the denominator of the integrand. This, along with one factor of $k_1$
or $k_2$ in the numerator, when symmetrised with respect to $k_i \to -
k_i$ again gives a factor of $c_{12}$ in the numerator, and thus renders
this integral IR finite. 

This analysis can be extended to the case where we have a ladder of $m$
photons, with the outermost photon mementum, $k_1$, controlling the
divergence, or where we have $m$ crossed photons, each of which must
vanish before a divergence arises with respect to terms containing one
power of $k_i, i \in [1, m]$; in each case, the symmetry under $k_i \to
- k_i$ removes the divergence, leaving behind an IR finite result.

We now consider the case when not all the intermediate photon
vertices correspond to $T \neq 0$ photons; i.e., they do not come with 
$\delta (k_z^2)$ in their propagators. If the $T = 0$ photon belongs to
the controlling set, then any factors of $k_z$ in the numerator anyway
make the integral finite (as the $T = 0$ terms are only logarithmically
divergent). If not, the $T = 0$ photons are symmetric under $k_z \to -
k_z$ as well, and so the previous analysis for PkN's goes through without
any problem here as well.

Finally, we consider the case where we have mixed $K$ and $G$ type
photons in the matrix element when we insert the $(n+1)$-th $G$ photon.
Every $K$ photon reduces to an overall factor of $B$ multiplying the
rest of the matrix element, as shown in the previous subsection. This
reduction did not depend on whether there were $K$ or $G$ photons at the
other vertices. What we will have left over will then be only the $G$
photon contribution; our above analysis then goes through without a
hitch.

Hence we have shown that the $G$ photon separation results in an IR
finite answer for the entire finite temperature contribution, when all
the photons in the set are virtual photons. This result also holds when
one of the intermediate photon vertices corresponds to a real photon
emission/absorption. The leading divergences again cancel between the
$g_{\mu\nu}$ and $b k_\mu k_\nu$ terms (this analysis did not depend on
whether the intermediate vertices had real or virtual photons
attached). Hence, we need examine only the one PkN terms. These survive
only at $T \neq 0$. Since real photon emission comes with a factor
$\theta(k_r^0)$, it is not symmetric with respect to $( k_r \to - k_r)$,
and it seems as though the symmetry procedure described above fails,
leaving logarithmic subdivergences. However, this is not so; the rule is
now to symmetrise the integrand only with respect to the remaining
(virtual) photon momenta. Such a procedure then gives rise to terms
with $c_{ir}$ or $a_i$ in the numerator, where $k_i$ is one of the
controlling photons. Hence these contributions are IR finite as well. 

We have thus shown that all skeletal graphs are IR finite with respect
to insertion of $G$ photons. We show that this result
continues to hold when we ``flesh'' out the skeletal graphs to obtain
the original $n$-photon vertex graph by adding self energy or vertex
parts, in Appendix A. In fact, the arguments set forth by GY are
equally valid here, as the temperature
does not affect the power counting when these insertions are made. 

We summarise the results of this section, by stating that only $K$
photons contain IR divergences; every $K$ photon contribution factorises
out of the expression for the matrix element as a factor, $B$ (defined
in eq. (3)). Every $G$ photon contribution is IR finite. 

\subsubsection{Matrix Element for virtual photons}

In general, a graph may contain $n_K$ virtual $K$ photons and $n_G$
virtual $G$ photons, $n = n_K + n_G$ being the total number of photon
vertices.
Each distinct graph can arise in $n!/n_K! n_G!$ ways so that the
matrix element corresponding to $n$ virtual photons is a sum of all
such possible contributions:
$$
\begin{array}{rcl}
\displaystyle \sum_{n = 0}^\infty \by{1}{n!} {\cal M}^n & = &
          \displaystyle \sum_{n = 0}^\infty \sum_{n_K = 0}^n
	  \by{1}{n_K!} \by{1}{(n - n_K)!} M_{n_G, n_K} \\
& = & \displaystyle \sum_{n_K = 0}^\infty \sum_{n_G = 0}^\infty
          \by{1}{n_K!} \by{1}{n_G!} M_{n_G, n_K}~,
\end{array}
\eqno(19) 
$$
where the individual matrix element has just been shown to contain a
factor $B$ for each $K$ photon times an IR finite contribution from
the $G$ photons:
$$
M_{n_G, n_K} = ({B})^{n_K} \, M_{n_G,0} \equiv (B)^{n_K} \, M_{n_G}~,
\eqno(20)
$$
where $B$ is given in eq. (3). Hence, the total matrix element can be
written as,
$$
\begin{array}{rcl}
\displaystyle \sum_{n = 0}^\infty \by{1}{n!} {\cal M}^n & = &
\displaystyle \sum_{n_K = 0}^\infty \by{({B})^{n_K}}{n_K!}
\sum_{n_G = 0}^\infty
\by{1}{n_G!} M_{n_G}~, \\
& = & \displaystyle {\rm e}^B \sum_{n = 0}^\infty \by{1}{n!} M_{n}~, 
\end{array}
\eqno(21)
$$
where the $M_n$ are now IR finite and the IR divergent part has been
explicitly projected out in exponential form. 

\subsection{Real photon emission/absorption}

The case of emission and absorption can be treated uniformly; the phase
space element in eq. (10) maintains the correct statistical factor in each
case. Here, it is the polarisation sum in the cross section, 
$$
\sum_{\rm pol} \epsilon_\mu \epsilon_\nu = - g_{\mu\nu}~,
\eqno(22) 
$$
that is separated into a $K$ and $G$ piece:
$$
\begin{array}{rcl}
- g_{\mu\nu} & = & - (g_{\mu\nu} - \tilde{b} k_\mu k_\nu) -
\tilde{b} k_\mu k_\nu \\
& = & - \widetilde{G}_{\mu\nu} - \widetilde{K}_{\mu\nu}~,
\end{array}
\eqno(23)
$$
with $\tilde{b}$ defined as 
$$
\tilde{b}(p_a, p_b) = \by{p_a \cdot p_b }{k \cdot p_a k \cdot p_b}~,
\eqno(24) 
$$
where $p_b$ ($p_a$) is $p$ or $p'$ according to whether the emission or
absorption takes place on the $p$ or $p'$ leg in the matrix element, $M$
(its conjugate $M^\dagger$). Now the calculation proceeds exactly as in
the virtual case. It turns out (see Appendix B for details) that the
insertion of a $\widetilde{K}$
photon into a graph with $n$ photon vertices results in an overall factor
multiplying the {\it cross section}, given by (neglecting the phase
space factors for the sake of clarity),
$$
\begin{array}{rcl}
 & \sim & - e^2 ( \widetilde{b} (p, p) +
 \widetilde{b} (p', p') - 2 \widetilde{b} (p, p')) \\
 & = & -e^2 J^2(k)~.
\end{array}
\eqno(25) 
$$
This contains both logarithmic as well as linearly divergences in the IR
of $k$, just as the corresponding factor which factored out of the
{\it matrix element} for virtual photons. Hence the IR divergences in
the real photon cross section factorise as well, although in terms of
$J^2$ (defined in eq. (2)) rather than $\widetilde{J}^2$.

The contributions of the $\widetilde{G}$ photons are again IR finite, as
in the virtual case, as can be seen by applying the identity analogous
to eq. (16),
$$
p_\mu^a p_\nu^b \widetilde{G}_{\mu\nu} = 0~.
\eqno(26)
$$
Note that this is exact with no corrections of ${\cal O}(k)$ since 
$\delta(k^2)$ always applies for the entire real photon contribution,
not only for the $T \neq 0$ piece. That is, the leading divergences
cancel once again between the $g_{\mu\nu}$ and $\tilde{b} k_\mu k_\nu$
parts of $\widetilde{G}$. Once again, we have to worry about one PkN terms
only for $T \neq 0$ parts of the cross section. While the real photon
cross section is certainly not, in general, symmetric under $k \to - k$
(because of the theta function constraint on the energy component of
$k$), we see that the $T \neq 0$ part of the cross section is symmetric
(since it includes both soft photon emission and absorption). We therefore
retain only $T \neq 0$ parts of the phase space element in our analysis
of the one PkN terms. We then symmetrise the resulting integrand with
respect to $k \to - k$ and obtain a result similar to that obtained for
the virtual photons, viz., the sub-leading logarithmic divergences are
removed and the one PkN terms are also IR finite. 

We have to now consider what happens when one of the intermediate
photons, $k_l$, contributes as a $T = 0$ photon, i.e., its corresponding
momentum cannot be flipped in the matrix element to perform the
symmetrisation which leads to a finite result. If this photon is not a
controlling photon, it is irrelevant (there is no divergence associated
with the vanishing of this momentum) and the analysis holds. If it is a
controlling photon, then the sub-divergence occurs only when all such
controlling momenta vanish; however, any power of $k_l$ in the numerator
renders that integral finite, since $k_l$ is a $T = 0 $ photon and so
the one PkN terms do not contribute. This analysis can be extended to
the case where more than one intermediate photon contributes through
its $T = 0$ part. 

Finally, if some of the intermediate $k_i$ correspond to virtual
photons, this will also not upset the analysis, since the virtual photon
momenta can always be flipped (the result is always symmetric in the
loop momentum). 

For more details, we again refer the reader to the paper by GY; the
extension of their arguments to the finite temperature case at hand using
the symmetry arguments set forth in the previous section is straightforward. 

\subsection{The cross section to all orders}

There is a slight difference between the real and virtual
calculations since the real photon insertion also changes the energy
momentum conservation relation (since the momentum $k$ is physically
lost or gained). Since a given combination of $n_K$ $\widetilde{K}$
photons and $n_G$ $\widetilde{G}$ photons can occur in $n!/n_K! n_G!$
ways, the cross section for $n$ real photon emission/absorption is
$$
\begin{array}{rcl}
\d\sigma_n^{\rm real} & = & \displaystyle \sum_{n_k = 0}^n \int
\prod_{i = 1}^{n_K}
\d\phi_{k_i} [-e^2 J^2] \by{1}{n_G!} \prod_{j = n_K+1}^n \d\phi_{k_j}
[-\widetilde{G}_{\mu\nu} \vert M_{n_G} \vert^2_{\mu\nu}] \times \\
 & & \hspace{4cm} \displaystyle (2\pi)^4 \delta^4(q + p - p' -
 \sum_{l = 1}^n k_l)~.
\end{array}
\eqno(27)
$$
Disentangling the $k$ dependence in the energy--momentum conserving
delta function using
$$
(2\pi)^4 \delta^4 (q + p - p' - \sum_l k_l) = \int \d^4 x
\exp(-i(q + p - p') \cdot x ) \prod_l \exp (i k_l \cdot x)~,
\eqno(28) 
$$
it will be possible to factorise and exponentiate the IR divergent piece
coming from real photons as well. On including the multiplicative
virtual contribution,
$$
\vert Z \vert^2 = \exp(B + B^*)~,
\eqno(29) 
$$
we get the total cross section at this order to be, 
$$
\begin{array}{rcl}
\d\sigma_n & = & \displaystyle \int \d^4 x \;{\rm e}^{i(q + p - p')\cdot x}
\;{\rm e}^{(B  + B^*)} {\rm e}^{\hat{B}} \sum_{j = 0}^\infty \by{1}{j!}
\prod_{j = 0}^n \int \d\phi_{k_j} {\rm e}^{i k_j \cdot x}
\left[ -\widetilde{G}^{\mu_j \nu_j})  \{M^\dagger_{n_G}\}_{\mu_j}
\{m_{n_G}\}_{\nu_j} \right] \\
 & \equiv & \displaystyle \int \d^4 x \;{\rm e}^{i(q + p - p')\cdot x} \;
{\rm e}^{(B + B^* + \hat{B})} \sigma^{\rm finite} (x)~,
\end{array}
\eqno(30) 
$$
where $\sigma^{\rm finite}$ contains the finite $G$ contributions of
both real and virtual photons and $\hat{B}$ is defined in eq. (5).
The exponentiated IR divergent pieces
$({B} + {B}^*)$ from virtual photon graphs and
$\hat{B}$ from real photon graphs add to an IR finite sum, as can be
seen by looking at their small $k$ behaviour, when $J^2$ reduces to
$\widetilde{J}^2$: 
$$
(B + B^*) + \hat{B} = e^2 \int \widetilde{\d\phi_k} \left[ \left\{J^2
(1 + 2N) \right\} - \widetilde{J}^2 \left\{(1+N) {\rm e}^{i k \cdot x} +
N {\rm e}^{ - i k \cdot x} \right\} \right]~,
\eqno(31)
$$
which was the result also obtained in the eikonal approximation by
Gupta et al. \cite{Gupta} with $J^2 = \widetilde{J}^2$. 

\section{Inclusion of Fermions}

Due to the abelian nature of QED, electronic corrections to the hard
scattering process occur in the form of closed fermion loops. By
itself, inclusion of electron loops does not affect the IR
behaviour of the hard scattering cross section. This is because the $T =
0$ contribution is IR finite \cite{IZ} and the inclusion of the finite
temperature piece does not change the IR behaviour because of the
small-$k$ form of the Fermi-Dirac distribution function (see eq. (8)).
However, we need to examine whether they induce divergences in the
photons they couple to. That is,
consider a photon of momentum $k_i$ entering a loop. We wish to compute
the loop contribution, $\omega(k_i, \cdots)$, when $k_i \to 0$. 

An electron loop with $m$-photon vertices can be thought of as an
$m$-point current current correlator (Green's function). Then, gauge
invariance implies that $k_{i,\mu} \omega^{\mu\ldots} = 0$. For
instance, the
$T =0$ 2-point function (photon self energy diagram) corresponding to 
a photon of momentum $k$, due to current conservation, can be expressed
as the purely transverse combination, 
$$
\omega(k) = (k^2 g_{\mu\nu} - k_\mu k_\nu) \overline{\omega}~,
$$
and strongly vanishes as $k \to 0$. Hence there can be no IR divergences
associated with such photons (the matrix element vanishes as the photon
momentum goes to zero). In general, if there are more photons entering
the loop, current conservation yields $k_\mu^i {\cal M}^{\alpha \beta
\ldots \mu \ldots} = 0$ for every photon $k_\mu^i$. In other words, the
contribution of a loop into which a scalar photon is inserted vanishes,
leaving behind only the transverse contribution, which is proportional
to powers of the external momenta \cite{Sterman}. This result also holds
at finite temperature, since it is a consequence of gauge
invariance \footnote{See the discussion in the next subsection, leading
to eq. (32) and below. A covariant analysis of the photon
self energy diagram is given in Ref. \cite{Weldonold}.}.


So far, we have considered the case where the external momentum enters
the electron line at a single vertex, $q$. This led to an unambiguous
separation of the electron line into $p$ and $p'$ legs. This need not
remain so when we have more than one hard scattering vertex (as in
Compton scattering) or when the external hard photon couples through a
closed fermion loop. In the latter case, the loop may attach to the
electron line through more than one virtual photon, with momenta $q_i, i
= 1, m$, $\sum_i^m q_i = q$. This leads to an ambiguity in defining the
$p$ and $p'$ legs uniqely, as it will depend on which of the $q_i$ is
used as the separating vertex. However, the separation of the IR divergence
is insensitive to the choice of $q_i$. This is because, as just shown,
the loop contribution vanishes if any of the $q_i$ vanishes. Hence all
the $q_i$ are non-zero, and there is a hard momentum flowing through the
entire part of the electron leg between $q_1$ and $q_m$ so that none of
the in-between propagators can vanish. There is thus no IR divergence
associated with the momenta flowing through any of these vertices; any
one of them can be chosen to separate the $p$ and $p'$ legs. The IR
factorisation then goes through, although the IR-finite contribution of
a given graph will depend on the choice of the `$q$' vertex. The proof
technique was also extended by GY to deal with graphs involving more
than one external electron. There appears to be no difficulty, in
principle, in applying these extensions to the finite temperature case;
here, however, we do not pursue this further. 

Finally GY showed, for the $T = 0$ case, that their technique (separation
into $G$ and $K$ photons) does not affect the usual analysis of
UV divergences. At $T \neq 0$, the Bose-Einstein (or
Fermi-Dirac) distribution function acts as a natural UV cut-off,
rendering the cross section UV-finite. 

We briefly address the question of thermal ``electron legs''. So far,
the external (incoming and outgoing) electrons and the hard photon were
not thermalised with respect to the heat bath. While we do not wish the
hard photon to be thermalised (we have assumed $-q^2 \gg T^2$), it is
interesting to consider the case when the external electron, with which
the hard photon scatters, is either a thermal electron, or gets
thermalised in the heat bath long before the hard scattering occurs (at
the vertex $q$). This is certainly possible since $p^2 = m^2$ which
need not be larger than $T^2$. In this case, the thermal electrons not
only occur as loop corrections in the relevant graphs, but also on the
$p$ and $p'$ legs\footnote{It is unlikely that the electron is
thermalised {\it after} the hard scattering. However, the analysis we
will present is applicable independently to the $p$ and $p'$
legs; hence we address the general case where both the legs have thermal
fermions. The thermal electron corrections can then be made on the
appropriate legs depending on the kinematics of the scattering.}.

Thermal electrons add a level of complexity to the computations
of the photon corrections; this is because of field doubling.
We have therefore to repeat the calculations of the previous
section, retaining matrix forms for the propagators of both electrons
and photons, as given in Section 2. We have already seen that there is
no new divergence associated with photon insertions on thermal fermion
loops. 

We first consider the case where the photon insertions (either real or
virtual) are made on the $p$ or $p'$ legs (and not on electron loops) in
an $n$-photon graph. 

\paragraph{Virtual Photon insertion on electron legs}: By electron legs,
we mean non-loop electron propagators. The vertices where the virtual
photons are inserted can be of either type-1 or type-2 since the
electron propagators are now thermal as well. However, the separation
of the virtual contribution
into $K$ and $G$ photons is unaffected by this increase in the number of
field types. This is because every photon propagator still comes with a
factor of $g_{\mu\nu}$, enabling the GY separation, as was done in the
earlier analysis with only (11) type photons and $T = 0$ fermions. We
therefore insert the photon in all possible ways into the graph,
remembering that the external legs must correspond to type-1 fields.
Consider the insertion of a $K$ photon. 

At any one vertex of insertion, we have the vertex factor $\sim
\not{\!k}$, sandwiched between two fermion propagators. We then use the
following generalised Feynman identity (cf. eq. (13)):
$$
S_{ay}(p - k) \not{\!k} S_{xa}(p) = i (-1)^{a+1} \left[ 
S_{xy}(p - k) \delta_{xa} - S_{xy}(p) \delta_{ya} \right]~.
\eqno(32)
$$
Here, the photon vertex of type-$a$ is inserted between the two vertices
of type $x$ and $y$ respectively. Hence, the result is again a
difference of two terms; there is again a pair-wise cancellation, giving
the same result as in the earlier analysis but with extra factors. This
is gratifying, also because the result that closed electron loops only
couple to transverse photons depends on such a cancellation. For
example, the $b_k (p, p)$ contribution is (writing $k_{n+1} \equiv k$
as usual), 
$$
{\cal M}^{n+1} \sim i e^2 \int \by{\d^4 k}{(2\pi)^4} b_k(p, p)
\delta_{qa} \delta_{qb} D_{ab} (k) {\cal M}^n~.
$$
Here $q$ corresponds to the vertex type of the hard vertex (where the
hard photon of momentum $q$ enters); there should be no confusion
between the index $q$ which denotes the vertex type of the hard photon
vertex and the momentum $q$ which is the momentum of the hard photon.
Since $q$ is an external non-thermal field, the corresponding vertex
must be of
type-1. Hence, $q = 1$, or $a = b = 1$ as well, from the delta
functions. Since only the (11) contribution is non-vanishing, the result
reduces to that of the previous case. This is true for the other
insertions, on the $p'$-leg, etc., as well. We again find that the $b_k
(p', p')$ contribution is zero, because of the disallowed diagram with a
self energy insertion on the outgoing electron line. We again symmetrise the
result between $b_k(p, p)$ and $b_k(p', p')$. In short, putting the
results together, the matrix element for virtual photon insertion turns
out to be an overall factor multiplying the original $n$-photon matrix
element, given by, (cf. eq. (14)),
$$
\begin{array}{rcl}
 & = & \by{ie^2}{2} \displaystyle \int \by{\d^4 k}{(2\pi)^4}
 \left[ b_k(p, p) + b_k(p', p') - 2 b_k(p, p') \right] \delta_{qa}
 \delta_{qb} D_{ab} (k) ~, \\
& = & \by{ie^2}{2} \displaystyle \int \by{\d^4 k}{(2\pi)^4}
     \left[ J^2 \right] D_{11} (k)~, \\
& \equiv & B~, 
\end{array}
\eqno(33)
$$
as before. Note that this analysis also includes consideration of
diagrams which are disallowed at lower order (with type-2 vertices next
to the external legs) and which are in fact found not to contribute at
the next order. 

We now consider insertion of virtual $G$ photons. Every electron
propagator can be expressed in the form,
$$
S_{ab} (p - k) = \left[F^{-1}\right] (\not{\!p} - \not{\!k} + m)~,
\eqno(34)
$$
for every combination of $a$ and $b$. (See Appendix C for the exact
forms of $F$). In the earlier analysis, we had $T = 0$ electrons alone:
$$
S^{T=0}(p - k) = \by{i}{(p-k)^2 - m^2} (\not{\!p} - \not{\!k} + m)~.
\eqno(35)
$$
Comparing eqs. (38) and (39), we see that the numerators, which were
crucial in showing the IR finiteness of the $G$ photons, are not altered
by the thermalisation of the electrons. Hence, the analysis in the 
equations from (15) through (16) remains unchanged when we include thermal
electrons. The cancellation between the IR divergent terms of
$g_{\mu\nu}$ and $b k_\mu k_\nu$ therefore holds for the leading
divergences. We have therefore only to consider the one PkN terms at
$T \neq 0$ which may possibly give subdivergent logarithmic
contributions. These are finite with respect to the $T = 0$ part of
the photon propagator; hence we consider only the $T \neq 0$ part:
we see that all the four components have the same leading IR behaviour.
Also they are symmetric in $k \to -k$, as is required in the analysis
of the one PkN terms. 

We focus on the $F$'s, the ``denominators'', in skeleton graphs, since
the numerators are the same as before. The $T=0$ electron propagators
come as factors of $L_0(l) = 1/(l^2 - m^2)$ while the relevant portion
of all the $T \neq 0$ parts is $L_T(l) = \delta(l^2 - m^2)$. 
Hence the various contributions are (1) all $L_0$ type propagators, (2)
some $L_T$ and some $L_0$ types, and (3) all $L_T$ types. The first case is
what has been discussed in the previous section. For the second case,
we consider again the ladder graph of fig.\ \ref{fig2}. There are
four propagators
here, with $l = (p' - k_1)$, $(p' - k_1 - k_2)$, $(p - k_1)$, and $p -
k_1 - k_2$, where $k_1$ is the controlling photon momentum. The
$L_0$ and $L_T$-type terms are dimensionally equivalent; however, we
see that while the one PkN term,
$$
\int \d^4 k_1 N(\vert {\bf k_1} \vert) \delta(k_1^2) {k_1^\rho}
L_0(p' -k_1) L_0(p - k_1)~,
$$
is logarithmically divergent, the term with one of the $L_0$'s replaced by
an $L_T$:
$$
\int \d^4 k_1 N(\vert {\bf k_1} \vert) \delta(k_1^2) {k_1^\rho}
L_0(p' -k_1) L_T(p - k_1)~,
$$
is finite because of the delta function. Hence terms containing either
$L_T(p - k_1)$ or $L_T(p' - k_1)$ are IR finite with respect to one PkN
terms. 
We have only to consider the case when both the $(p - k_1)$ and $(p' -
k_1)$ terms occur as $L_0$ terms. If all the other terms are also of $L_0$
type, then it reduces to Case (1). If even one term is an $L_T$ term, then
it contains a delta function involving $k_1$, since it is a controlling
momentum. This delta function evaluates to a constraint on $k_1$, for
example, for the graph of fig.\ \ref{fig2} and $l = (p - k_1 - k_2)$, this leads to
$$
k_1^0 = {\bf k_1} = \by{k_2^2 - 2 p \cdot k_2}{2(p - k_2)
\cdot \hat{k}_1}~,
$$
where $\hat{k}_1 = (1, {\bf{n}})$ and ${\bf n}$ is the unit vector in the
direction of ${\bf k}_1$. 
Since $p_0 \neq {\bf p}$, this is constant for fixed $k_2$. Hence,
although the term integrates to a logarithm, there is no associated
divergence since $k_1 = 0$ is not allowed unless $k_2$ also vanishes. 

Finally, since Case (3) contains only $L_T$ type terms, it is finite as
well. The analysis is equally applicable to cross graphs (such as
fig.\ \ref{fig3}). Hence the $G$ photon insertion is IR finite with respect
to thermal electron legs as well. 

Finally, we can ``flesh'' out the skeleton as before---this will depend
only on the (leading) numerator behaviour. Since this remains unchaged,
the earlier analysis holds here also.

\paragraph{Real Photon Insertion on Electron Legs}: The same
cancellation also occurs in the case of real photon emission/absorption.
Here, we can only insert type-1 fields since these are physical photons.
We find the result to be the same as in our previous analysis for real
photon insertion, with an additional factor of $\delta_{qa}$, where $a$
is the inserted photon type. Since $a$ must be 1, and so is $q$, the
delta function is always satisfied, and we get the same result as
before:
$$
\begin{array}{rcl}
\sigma^{n+1} & \sim & {{\cal M}^{n+1}}^\dagger {\cal M}^{n+1}
(-\widetilde{b}) \d \phi_k~, \\
& = & -e^2 \displaystyle \int \d \phi_k \left[ \widetilde{b}_k (p, p) + 
\widetilde{b}_k (p', p') - 2 \widetilde{b}_k (p, p') \right]
{{\cal M}^n}^\dagger {\cal M}^n~.
\end{array}
\eqno(36)
$$
Several of the identities used in obtaining these results are given in
Appendix C. The factorisation of the $\widetilde{K}$ photon contribution
therefore goes through in both virtual and real photon insertions on the
$p$ and $p'$ electron legs. 

An analysis similar to that used for $G$ photon insertion can be made to
show that $\widetilde{G}$ photons insertions yield IR finite
contributions, as before. 

Hence, the separation of $K$ and $G$ photon contributions, which are IR
divergent and IR finite respectively, can be made even when the hard
scattering occurs off thermal fermions. The IR divergences cancel as
usual between real and virtual diagrams, order by order, at every order.
The Bloch Nordsieck theorem is therefore valid here as well. 

\section{Conclusions}
Weldon showed the cancellation of IR divergences in a hard scattering
process occurring in a heat bath with thermal photons, within
the eikonal (or soft photon) approximation. We have now extended his
proof to show the IR finiteness of the exact cross sections for hard
processes with one electron each in the initial and
final state, and arbitrary number of real or virtual photons. We then
introduced electrons into the heat bath, and showed that the full
thermal QED hard scattering cross section is IR finite as well. We also
note that the analysis made in \cite{Gupta} on correlations between
final state momenta are therefore valid, even without resorting to
the eikonal approximation. 

We have used the technique of Grammer and Yennie (GY), developed for the
zero temperature case, to accomplish the factorisation and cancellation
of the IR divergent terms. Sets of diagrams then factor neatly,
with no regrouping between real and virtual diagrams being required in
order to effect this factorisation. The IR divergent parts can be
factorised at arbitrary order, and hence exponentiated,
and cancel between real and virtual sets of diagrams. The procedure also
prescribes a method to obtain the IR finite parts of such cross sections. 
For comments on the general utility and applicability of the GY technique
in various calculations, the reader is referred to the original paper of
GY \cite{GY}. IR finiteness to all orders is thus easily demonstrated
using this technique. We comment that, in this calculation of hard
scattering in a heat bath of electrons and photons, the thermal
modifications (of the propagators and of the phase space)
are simple enough that the original proof of Grammer and Yennie, of IR
finiteness of QED at zero temperature, can be easily generalised to
include the $T \neq 0$ case. This is because the hard scattering scale
is taken to be much larger than the temperature, $-q^2 \gg T^2$, so
that thermal effects are relatively soft. An increase in temperature
(from ``warm'' to ``hot'' QED) would result in a need to resum hard
thermal loops a la Braaten and Pisarski \cite{BP}. The question then
is whether the IR finiteness would still hold in such a case. This is
definitely worth studying, especially in view of recent results in the
field \cite{Blaizot}, although outside the scope of the present work.

\paragraph{Acknowledgements:} I thank V. Ravindran who was responsible
for much of the interest in, and understanding of the initial stages of
this work. I am thankful to  Profs. G. Sterman and P. van Nieuwenhuizen
for many interesting discussions. I also thank Rahul Basu for several
clarifications, and Prof. E. Reya for constant encouragement and support. 

\vspace{1cm}

\noindent {\large \bf {Appendix A}}

\paragraph{The $K$ photon insertions}: 
We present details of the results for the factorisation of the IR
divergent terms from graphs with an additional virtual $K$ photon
insertion. We specifically consider a photon insertion across the
external hard photon vertex, i.e., the extra virtual photon connects the
initial electron leg to the final electron leg.

Consider the graph with (symmetrised) $n$ photon vertices shown in
fig.\ \ref{fig1}.
There are $r$ vertices on the $p$ leg of the graph (everything before
the interaction with the external hard photon, $q$), numbered from $1$
to $r$, and $s$ vertices on the $p'$ leg (the electron line to the right
of the hard photon vertex), numbered from $s$ to $1$; $(r + s) = n$. The
momentum corresponding to the electron line to the right of the $m$-th
vertex on the $p$ leg is therefore $(p - \sum_{i=1}^m k_i)$ while the
momentum corresponding to the electron line to the left of the $m$-th
vertex on the $p'$ leg is $(p' + \sum_{i=1}^m l_i)$. We will denote this
in short by $(p - \sum_m k_i)$ and $(p' + \sum_m l_i)$ respectively. We
now insert a virtual photon with momentum $k_{n+1}$ and examine the IR
behaviour of the resulting graph.
The $(n+1)$-th photon must connect from the
'left' to the 'right' of the vertex $q$, which corresponds to the
interaction with the hard external photon of momentum $q$ (in order for
them all to have the same overall $b(p, p')$ factor). We study the
insertions case by case, beginning with the case when the final point
insertion is on the $p'$ leg.

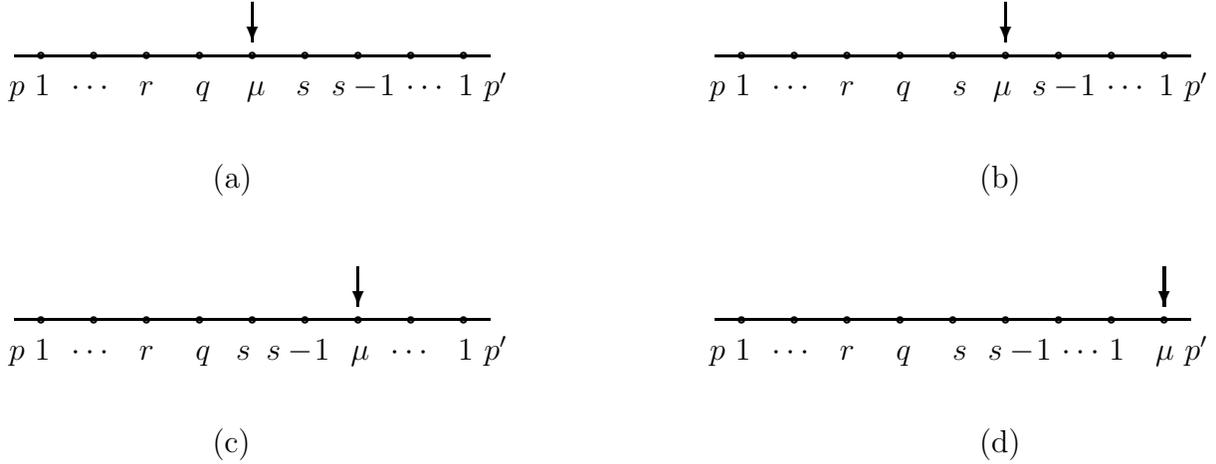
\begin{figure}[ht]

\noindent \begin{picture}(380,200)(0,0)
\thicklines

\put(95,170){\vector(0,-1){15}}
\put(5,150){\line(1,0){180}}
\multiput(15,150)(20,0){9}{\circle{2}}
\put(3,135){$p$ 1 ~$\cdots$ ~~$r$ \quad $q$ \quad $\!\mu$ ~ $s$
~$s-\!1$ $\cdots$ $\, 1$ $p'$}
\put(80,100){(a)}

\put(380,170){\vector(0,-1){15}}
\put(270,150){\line(1,0){180}}
\multiput(280,150)(20,0){9}{\circle{2}}
\put(268,135){$p$ 1 ~$\cdots$ ~~$r$ \quad $q$ \quad $s$ ~ $\!\mu$
~$s-\!1$ $\cdots$ $\, 1$ $p'$}

\put(370,100){(b)}

\put(135,70){\vector(0,-1){15}}
\put(5,50){\line(1,0){180}}
\multiput(15,50)(20,0){9}{\circle{2}}
\put(3,35){$p$ 1 ~$\cdots$ ~~$r$ \quad $q$ ~ $\!s$~ $\!s-\!1$
~$\mu$ ~$\cdots$ ~ 1 $p'$}
\put(80,0){(c)}

\put(440,70){\vector(0,-1){15}}
\put(270,50){\line(1,0){180}}
\multiput(280,50)(20,0){9}{\circle{2}}
\put(268,35){$p$ 1 ~$\cdots$ ~~$r$ \quad $q$ \quad $s$ ~$s-\!1$
$\cdots$ $1$ ~ $\mu$ $p'$}

\put(370,0){(d)}

\end{picture}

\caption[dummy]{\small This demonstrates insertion of the $(n+1)$-th
virtual $K$ photon on the $p'$ leg in all possible ways.}
\label{fig4}

\end{figure}

\vspace{0.3cm}

\paragraph{Case 1}: The initial point is fixed on the $p$ leg. We must
then sum the contributions of the diagrams where the final point is just
to the right of the vertex labelled '$q$', to the right of the
vertex labelled 's', to the right of the vertex '$(s-1)$'
and so on, until it is to the right of the vertex '1' on the $p'$ leg.
Some of the graphs are shown in fig.\ \ref{fig4}. In each of these
terms, we must
make the substitution, in the photon propagator,
$$
g^{\mu\nu} \to b_{k_{n+1}} (p, p') k_{n+1}^\mu k_{n+1}^\nu~,
$$
since we are only considering the $K$ photons here. This means that
there is a factor, $\not{\!k}_{n+1}$ at the vertices where the $K$
photon is inserted. We then make use of the identities given in eq. (13).
The part of the matrix element corresponding to the $p'$-leg (that is,
to the right of the vertex '$q$') of figs.\ 4a and 4b then is, 
$$
\begin{array}{rcl}
{\cal M}^{4a} & \sim & \overline{u}_{p'} \gamma_1 \ldots
\gamma_{s-1} \bynd{\displaystyle 1}{{\displaystyle \not{\!p}'} +
\sum_{\scriptstyle s-1} {\displaystyle \not{\!\ell}_i - m}} \gamma_s\!
\left[\bynd{\displaystyle 1}{{\displaystyle \not{\!p}'} +
\sum_{\scriptstyle s} {\displaystyle \not{\!\ell}_i -
\!\!\not{\!k}_{n+1} - \!m}} - \bynd{\displaystyle 1}{{\displaystyle \not{\!p}'}
+ \sum_{\scriptstyle s} {\displaystyle \not{\!\ell}_i - \!m}} \right]\!
(-ie\!\!\not{\!\epsilon}_q) \ldots \\
 & \equiv & M'_2 - M'_1~, \\
{\cal M}^{4b} & \sim & \overline{u}_{p'} \gamma_1 \ldots
\gamma_{s-1} \!\left[ \bynd{\displaystyle 1}{{\displaystyle \not{\!p}'}
+ \sum_{\scriptstyle s-1} {\displaystyle \not{\!\ell}_i -
\not{\!k}_{n+1} - m}} - \bynd{\displaystyle 1}{{\displaystyle \not{\!p}'}
+ \sum_{\scriptstyle s-1} {\displaystyle \not{\!\ell}_i - m}}
\right] \!\gamma_s \times \\
& & \qquad \qquad \qquad \qquad \qquad \qquad \qquad \qquad \qquad
\qquad \!\quad
\bynd{\displaystyle 1}{{\displaystyle \not{\!p}'} + \sum_{\scriptstyle s}
{\displaystyle \not{\!\ell}_i - k_{n+1} - m}} 
(-ie\!\!\not{\!\epsilon}_q) \ldots \\ 
 & \equiv & M'_4 - M'_3~; ~~\mbox{similarly}, \\
{\cal M}^{4c} & \sim & M'_6 - M'_5~. 
\end{array}
$$
The ellipses to the right of $\not{\!\epsilon}_q$ indicate the
$p$-leg part of the matrix element which
is held fixed while the final point insertion is made at all possible
points on the $p'$-leg. 
We see that the second term in the expression for fig.\ 4b cancels with
the first term for fig.\ 4a, i.e., $M'_3 = M'_2$. The second term of
fig.\ 4c will similarly cancel with the first term of fig.\ 4b, i.e., $M'_5
= M'_4$, etc., thus giving rise to the cancellations mentioned in the
text. Finally, the last diagram of this set (fig.\ 4d) contributes,
$$
\begin{array}{rcl}
{\cal M}^{4d} & \sim & \overline{u}_{p'} \left[\not{\!k}_{n+1}
\by{1}{\not{\!p}' - k_{n+1} - m} \gamma_1 \ldots \right] 
(-ie\!\!\not{\!\epsilon}_q) \ldots  \\
 & \equiv & M'_{2s+1}~.
\end{array}
$$
Using the relations, 
$$
\begin{array}{rcl}
\by{1}{\not{\!p} - \not{\!k}_{n+1} - m} \not{\!k}_{n+1} u_p & = &
- u_p~, \\
\overline{u}_{p'} \not{\!k}_{n+1} \by{1}{\not{\!p}' - \not{\!k}_{n+1}
- m} & = & - \overline{u}_{p'}~,
\end{array}
\eqno(A.1)
$$
the total matrix element due to all diagrams corresponding to fig.\ 4, for
a fixed initial point on the $p$-leg, after pairwise cancellation, is
$$
\begin{array}{rcl}
{\cal M}^{p'} & = & (M'_2 - M'_1) + (M'_4 - M'_3) + \ldots + (M'_{2s} -
M'_{2s -1}) + M'_{2s+1}~, \\
& = & - M'_1~, \\
& \sim & -\overline{u}_{p'} \gamma_1 \by{1}{\not{\!p}' + \not{\!\ell}_1 - m}
\gamma_2 \ldots \gamma_s \by{1}{\not{\!p}' + \sum_s \not{\!\ell}_i - m}
(-ie\!\!\not{\!\epsilon}_q) \ldots u_p~. 
\end{array}
$$
Thus there is only one term (one of the terms from fig.\ \ref{fig4}a)
left over
after the cancellation. We see that there is no dependence on the
momentum, $k_{n+1}$ due to insertions on the $p'$ leg; in fact, what is
left is just the original $n$-photon matrix element. 

\paragraph{Case 2}: We now sum over all possible insertions of the
initial point on the $p$-leg. This can be inserted to the left of the
vertex labelled '1' on the $p$-leg in fig.\ 1, to the left of '2',
and so on. The relevant part ($p$-leg portion) of the matrix elements
for these contributions, with the final point fixed somewhere on the
$p'$-leg, (labelled with indices $l1$, $l2$, $\ldots$, $lq$), making use
of the identities, eq. (13), are, respectively,
$$
\begin{array}{rcl}
{\cal M}^{l1} & \sim & \ldots (- ie\!\!\not{\!\epsilon}_q)
\bynd{\displaystyle 1}{{\displaystyle \not{\!p}} - \sum_{\scriptstyle r}
{\displaystyle \!\not{\!k}_i - \!k_{n+1} \!- \!m}} \gamma_r \ldots \gamma_2
\bynd{\displaystyle 1}{{\displaystyle \not{\!p} - \!\not{\!k}_1 -
\!\not{\!k}_{n+1} - \!m}} \gamma_1 
\bynd{\displaystyle 1}{{\displaystyle \not{\!p} - \not{\!k}_{n+1} \!- \!m}}
\not{\!k}_{n+1} u_p\, ,\\
 & \equiv &  - M_1~, \\
{\cal M}^{l2} & \sim & \ldots (- ie\!\!\not{\!\epsilon}_q )
\bynd{\displaystyle 1}{{\displaystyle \not{\!p}} - \sum_{\scriptstyle r}
{\displaystyle \not{\!k}_i - \!k_{n+1} - m}} \gamma_r \ldots \gamma_2
\left[ \bynd{\displaystyle 1}{{\displaystyle \not{\!p} - \not{\!k}_1
- \!\not{\!k}_{n+1} - m}} - \bynd{\displaystyle 1}{{\displaystyle \not{\!p}
- \!\not{\!k}_1 - m}} \right] \gamma_1 u_p\, , \\
 & \equiv &  M_2 - M_3~, \\
{\cal M}^{lq} & \sim & \ldots (- ie\!\!\not{\!\epsilon}_q)
\left[\bynd{\displaystyle 1}{{\displaystyle \not{\!p}} -
\sum_{\scriptstyle r} {\displaystyle  \not{\!k}_i - \!\not{\!k}_{n+1} - m}}
- \bynd{\displaystyle 1}{{\displaystyle \not{\!p}} - \sum_{\scriptstyle r}
\! {\displaystyle \not{\!k}_i - \!m}} \right] \gamma_r 
\bynd{\displaystyle 1}{{\displaystyle \not{\!p}} -
\sum_{\scriptstyle r-1} {\displaystyle \not{\!k}_i - \!m}} \gamma_{r-1}
\ldots \gamma_1 u_p\, , \\
 & \equiv &  M_{2r} - M_{2r+1}~. 
\end{array}
$$
Using the identity shown in eq. (A.1), we find that $M_2 = M_1$, so there
is a similar pair-wise cancellation here as well. Proceeding as before,
and adding all the contributions, we obtain the total matrix element for
all possible insertions on the $p$-leg to be,
$$
\begin{array}{rcl}
{\cal M}^{p} & = & (-M_1) + (M_2 - M_3) + \ldots + (M_{2r} -
M_{2r+1})~, \\
 & = & -M_{2r+1}~, \\
& \sim & - \overline{u}_{p'} \ldots (-ie\not{\!\epsilon}_q)
\by{1}{\not{\!p} - \sum_r \not{\!k}_i - m} \gamma_r \ldots
\gamma_1 u_p~. 
\end{array}
$$
Again, there is only one term left over after pairwise cancellation;
again this is independent of $k_{n+1}$ and is effectively just the
original $n$-photon contribution. Combining cases (1) and (2), the
matrix element corresponding to insertion of a virtual $K$ photon at all
possible initial points (on the $p$-leg) and all possible final points
(on the $p'$ leg) is,
$$
{\cal M}^{n+1} \sim \overline{u}_{p'} \gamma_1 \ldots \gamma_s 
\by{1}{\not{\!p}' + \sum_s \not{\!\ell}_i - m}
(-ie\not{\!\epsilon}_q)
\by{1}{\not{\!p} - \sum_r \not{\!k}_i - m}
\gamma_r \ldots \gamma_1 u_p~, 
$$
and is hence proportional to the $n$-photon matrix element, ${\cal
M}^n$. Putting back the overall factors, including $(-b(p, p'))$ that we
have so far neglected, the $k_{n+1}$ dependence reduces to an overall
factor multiplying the $n$-photon matrix element, given by (writing
$k_{n+1} \equiv k$ for simplicity),
$$
{\cal M}^{n+1} = ie^2 \int \by{\d^4k}{(2\pi)^4} \left[ \by{1}{k^2 + i
\epsilon} - 2 \pi \delta(k^2) N (\vert {\bf{k}} \vert) \right] (-b_k (p,
p')) {\cal M}^n~.
\eqno(A.2)
$$
\paragraph{Case 3}: A similar analysis can be made for the case when
both initial and final points are on the $p$-leg. The insertions are
made in all possible ways; to avoid double counting, we insist
that the final point must always be to the right of (later than) the
initial point. The result (where again a pairwise cancellation occurs)
can be straightforwardly computed as before to get
$$
{\cal M}^{n+1} \sim (+ b_k (p, p)) {\cal M}^n~,
$$
where the overall factors are the same as in eq. (A.2); note the relative
sign difference with respect to (A.2). Again, the $k_{n+1}$ dependence
reduces to an overall factor multiplying the original $n$-photon result.

\paragraph{Case 4}: When all possible insertions are made on the
$p'$-leg alone, we have to disallow the diagram where the
inserted photon is a self energy insertion on the final electron line,
to compensate for wave function renormalisation. It then turns out
that the
$b(p', p')$ contribution vanishes. (The contribution of every set of
graphs for a fixed initial point and all possible final points is in
fact zero. The disallowed graph would have contributed a factor $\sim
b(p', p') {\cal M}^n$, symmetric with the $b(p, p)$ case; however, since
this graph is disallowed, there is no net $b(p', p')$ contribution).
Since the outermost self energy insertion could equally well have been
disallowed on the $p$-leg, we symmetrise the results of Cases (3) and
(4) by writing
$$
b(p, p) = \by{1}{2} \left[ b(p, p) + b(p', p')\right]~.
$$
Then the total $K$-photon contribution is an overall factor multiplying
the original $n$-photon matrix element:
$$
B = i e^2 \int \by{\d^4k}{(2\pi)^4} \left[ \by{1}{k^2 + i
\epsilon} - 2 \pi \delta(k^2) N (\vert {\bf{k}} \vert) \right]
\by{1}{2} \left[b_k (p, p) + b_k (p', p') - b_k (p, p') \right]~,
\eqno(A.3)
$$
as obtained in eq. (14).

\paragraph{The $G$ photon insertions}: We have seen that the skeleton
graph is IR finite with respect to insertion of virtual $G$ photons. We
have to check that this result is not spoiled by self energy and vertex
corrections. The arguments are essentially those originally put forward
by GY, and are merely presented here for completeness. As can be seen,
finite temperature does not change the validity of these arguments.

When we make a self energy insertion, the inserted photon
momentum, $k_l$, cannot be part of the controlling set (by definition).
Hence there cannot be a divergence associated with the vanishing of
$k_l$. However, we have to check that the degree of divergence is not
worsened with respect to any momentum in the controlling set. Let the
self energy insertion be made to an electron which had momentum $p -
k_c$ flowing through it; $k_c$ is a combination of the controlling
momenta. The insertion of $k_l$ thus replaces the propagator,
$1/(\not{\!p} - \not{\!k}_c - m)$ by
$$
\by{1}{(\not{\!p} - \not{\!k}_c - m)} \gamma_\mu 
\by{1}{(\not{\!p} - \not{\!k}_c - \not{\!k}_l - m)} \gamma_\nu 
\by{1}{(\not{\!p} - \not{\!k}_c - m)}~ G_{k_l}^{\mu \nu},
\eqno(A.4)
$$
and so seems to add more factors of $k_c$ in the denominator. However, it
was shown by GY that the intermediate propagator (due to the self energy
insertion) has terms of the form,
$$
\gamma_\mu \by{1}{(\not{\!p} - \not{\!k}_c - \not{\!k}_l - m)}
\gamma_\nu G_{\mu\nu} \sim \by{A (\not{\!p} - m) + B \not{\!k}_c +
C p \cdot k_c + \ldots}{ -a_{cl} + c_{cl}}~,
$$
where the ellipses denote powers of $k_l$ in the numerator. We wish to
assure ourselves that the divergence is not worsened by the addition of
this $k_l$ photon. Accordingly, we take $k_c \to 0$, and use the mass-shell
condition for $p$, upon which the expression (A.4) reduces to, 
$$
\by{1}{\not{\!p} - k_c + m} \left\{ (B 2 p \cdot k_c  + m\, C \, 
2 p \cdot k_c)
\by{1}{-2 p\cdot k_l + k_l^2} + {\cal O} (k_c^2)
\right\} \by{1}{(p - k_c)^2 - m^2}~.
\eqno(A.5)
$$
Hence, the extra power of $k_c$ coming from the term $(p \cdot k_c)$ in
the numerator cancels that of $((p - k_c)^2 - m^2)$ in the denominator,
so that the degree of divergence with respect to the controlling photon
momenta is not changed on insertion of the self energy piece. A similar
analysis can be performed for vertex insertions. Let the insertions be
made at vertices $\alpha$ and $\beta$ respectively. If $k_c$ is the
combination of photon momenta flowing out of the vertex, say, $\alpha$,
then the original factor,
$$
\by{1}{(\not{\!p}_r - m)} \gamma_\alpha 
\by{1}{(\not{\!p}_r - \not{\!k}_c - m)}~,
\eqno(A.6)
$$
where $p_r$ is the electron momentum to the right of the vertex, and
$p_r - k_c$ to the left of it, is replaced by
$$
\by{1}{(\not{\!p}_r - m)} \left\{ 
\gamma_\mu \by{1}{(\not{\!p}_r - \not{\!k}_l - m)}
\gamma_\alpha \by{1}{(\not{\!p}_r - \not{\!k}_c - \not{\!k}_l - m)}
\gamma_\nu  G_{k_l}^{\mu\nu} \right\} 
\by{1}{(\not{\!p}_r - \not{\!k}_c - m)}~.
\eqno(A.7)
$$
GY then analysed the extra contribution (shown in curly braces) \cite{GY}
dropping all powers of the controlling momenta $k_c$ in the numerator.
When $k_c \to 0$, $p_r \to p$ and either the extra contribution vanishes
because of the mass shell condition on $p$, or else it reduces to a
factor $p_\alpha$. Similarly, the contribution at the other vertex also
either vanishes, or reduces to $p_\beta$; hence the addition of the
vertex piece does not affect the original result (which also yielded an
overall factor $p_\alpha p_\beta$, as seen in the earlier section, where
the IR divergence cancelled due to the structure of $G$). Hence the
divergence with respect to the controlling photon momenta is not
worsened. When powers of $k_c$ are retained in the numerator, the result
is more convergent than the original one; hence all these terms give IR
finite contributions as well. Finally, when the vertex correction is
added just across the vertex where the hard photon (with momentum $q$)
enters, this will form a ladder graph where the newly inserted photon is
the innermost. Hence the divergence will depend on other controlling
($G$) photons which lie outside this insertion; these were already shown
to yield finite remainders; hence this insertion does not affect the
result as well. Hence the analysis is not affected by adding self energy
or vertex parts to the skeleton.

\vspace{1cm}

\noindent {\large \bf Appendix B}

\vspace{0.3cm}

\noindent A similar analysis can be done for insertions of
$\widetilde{K}$-type
real photons. It turns out that the insertion of a real photon with
momentum $k_{n+1}$ on the $p'$-leg in all possible ways into an
$n$-photon graph (with a factor $\not{\!k}_{n+1}$ at the insertion
vertex) results in a matrix element which reduces, on pairwise
cancellation of terms, to the $n$-photon matrix element:
$$
{\cal M}^{n+1} = (ie) {\cal M}^{n}~.
$$
When the insertion is on the $p$-leg, it reduces to
$$
{\cal M}^{n+1} = - (ie) {\cal M}^{n}~.
$$
\paragraph{Case 1}: The insertion is on the $p'$ leg in both ${\cal M}$
and ${\cal M}^\dagger$: the cross section ($\sigma \sim {\cal
M}^\dagger {\cal M}$) for this case is
$$
\sigma \sim {\cal M}^{n+1,\dagger} {\cal M}^{n+1} 
 =  e^2(-\tilde{b} (p', p')) {{\cal M}^n}^\dagger {\cal M}^n ~.
$$
\paragraph{Case 2}: The insertion is on the $p$ leg in both ${\cal M}$
and ${\cal M}^\dagger$: the cross section is
$$
\sigma \sim {\cal M}^{n+1,\dagger} {\cal M}^{n+1} 
 =  e^2(-\tilde{b} (p, p)) {{\cal M}^n}^\dagger {\cal M}^n ~.
$$
\paragraph{Case 3}: The insertion is on the $p$ ($p'$) leg in ${\cal M}$
(${\cal M}^\dagger$) and on the $p'$ ($p$) leg in ${\cal M}^\dagger$
(${\cal M}$): the cross section is
$$
\sigma \sim {\cal M}^{n+1,\dagger} {\cal M}^{n+1} 
 =   - e^2(-\tilde{b} (p, p') -  \tilde{b} (p', p) )
 {{\cal M}^n}^\dagger {\cal M}^n ~.
$$
Combining all possible insertions, the squared matrix element for the
insertion of a $\widetilde{K}$ real photon into an $n$-photon graph is
an overall $k_{n+1}$ dependent factor times the original result; this
factor (writing $k_{n +1} \equiv k$ as usual) is,
$$
\vert {\cal M}^{n+1} \vert^2 = -e^2 \left( \tilde{b}(p, p)
  + \tilde{b} (p', p') -  2 \tilde{b} (p, p')\right)~,
$$
which is eq. (25). On including the phase space factor for this extra
photon, and the $\exp(i k \cdot x)$ factor arising from fourier
transforming the energy--momentum conserving delta function, this
contribution evaluates to $\hat{B}$ as defined in eq. (5). 

\vspace{1cm}

\noindent {\large \bf Appendix C}

\vspace{0.3cm}

\noindent We give here the relevant identities used in the computation
with thermal fermions. 

\begin{enumerate}
\item The Electron Propagator:
$$
\begin{array}{rclrcl}
S_{11}(p) & = & F^{-1}_p (\not{\!p} + m)~; \\
S_{12}(p) & = & G^{-1}_p (\not{\!p} + m)~; \\
S_{12}(p) & = & -S_{21}(p)~;  \\
S_{22}(p) & = & {F_p^*}^{-1} (\not{\!p} + m)~;
\end{array}
$$
where
$$
\begin{array}{rcl}
F^{-1}_p  & = & i/(p^2 - m^2 + i \epsilon) + 2\pi N_f(\vec {\bf p}
\vert)~, \\
G^{-1}_p & = & F^{-1}_p \epsilon(p_0) \exp(\vert p_0 \vert/(2T))~, \\
F^{-1}_p (p^2 - m^2) & = & i~, \\
{F_p^*}^{-1} (p^2 - m^2) & = & -1~, \\
G^{-1}_p (p^2 - m^2) & = & 0~.
\end{array}
$$
\item The Generalised Feynman Identities:
$$
\begin{array}{rcl}
S_{\mu c}(p -k) \not{\!k} S_{b\mu} (p) & = & i (-1)^{\mu + 1} \left[ 
S_{b c}(p -k) \delta_{b\mu} - S_{b c}(p) \delta_{c\mu} \right]~; \\
S_{\mu c}(p ) \not{\!k} S_{b\mu} (p-k) & = & i (-1)^{\mu + 1} \left[ 
S_{b c}(p -k) \delta_{c\mu} - S_{b c}(p) \delta_{b\mu} \right]~; \\
S_{\mu c}(p'+k) \not{\!k} S_{b\mu} (p') & = & i (-1)^{\mu + 1} \left[ 
S_{b c}(p') \delta_{c\mu} - S_{b c}(p'+k) \delta_{b\mu} \right]~; \\
S_{\mu c}(p') \not{\!k} S_{b\mu} (p'+k) & = & i (-1)^{\mu + 1} \left[ 
S_{b c}(p') \delta_{b\mu} - S_{b c}(p'+k) \delta_{c\mu} \right]~; 
\end{array}
$$
\item Other Useful Formul\ae :
$$
\begin{array}{rcl}
S_{\mu a}(p -k) \not{\!k} u_{p} & = & -i (-1)^{\mu + 1} 
\delta_{a\mu} u_{p}~; \\
\overline{u}_{p'} \not{\!k} S_{a\mu}(p'+k)  & = & -i (-1)^{\mu + 1} 
\delta_{a\mu} \overline{u}_{p'}~. 
\end{array}
$$
\end{enumerate}


\begin{thebibliography}{99}

\bibitem{bkgnd} N.P. Landsman, Ch. G van Weert, Phys. Rep. {\bf 145},
141 (1987); A.J. Niemi and G.W. Semenoff, Ann. Phys. (N.Y.) {\bf 152},
105 (1984). 
\bibitem{qgp} See, for instance, M. Gyulassy and X.-N. Wang, Nucl. Phys.
{\bf B420}, 583 (1994); J.-C. Pan and C. Gale, Phys. Rev. {\bf D50},
3235 (1994); S. Gupta, Phys. Lett. {\bf B347}, 387 (1995). 
\bibitem{Gupta} S. Gupta, D. Indumathi, P. Mathews, and V. Ravindran,
Nucl. Phys. {\bf B458}, 189 (1996).
\bibitem{misc}See, for example, T.\ Altherr, P.\ Aurenche and
T.\ Becharrawy, Nucl.\ Phys.  {\bf B 315}, 436 (1989);
T.\ Grandou, M.\ LeBellac and D.\ Poizat, Phys.\ Lett. {\bf B 249}, 478
(1990) and Nucl.\ Phys. {\bf B 358}, 408 (1991);
T.\ Altherr, Phys.\ Lett. {\bf B 262}, 314 (1991);
P.\ V.\ Landshoff and M.\ Taylor, {Nucl.\ Phys.} {\bf B 430}, 683 (1994).
\bibitem{Weldon} H. A. Weldon, Phys. Rev. {\bf D49}, 1579 (1994). 
\bibitem{GY} G. Grammer, Jr., and D. R. Yennie, Phys. Rev. D8, 4332
(1973).
\bibitem{BN} F. Bloch and A. Nordsieck, Phys. Rev. 52, 54 (1937); see
also D.R. Yennie, S.C. Frautschi, and H. Suura, Ann. Phys. (NY) 13, 379
(1961). 
\bibitem{realtime} R.L. Kobes and G.W. Semenoff, Nucl. Phys. {\bf B260}, 714
(1985); A.J. Niemi and G.W. Semenoff, Nucl. Phys. {\bf B230}, 181
(1984); R. J. Rivers, see \cite{Rivers} below. 
\bibitem{Rivers} R.J. Rivers, {\it Path Integral Methods in Quantum
Field Theory}, R.\ J.\ Rivers, Cambridge University Press, Cambridge,
1987. 
\bibitem{Weldonold} H. A. Weldon, Phys. Rev. {\bf D26}, 1394 (1982). 
\bibitem{IZ} C. Itzykson and J.-B. Zuber, Quantum Field Theory, McGraw Hill
Publications, Singapore, 1985, pg 230; see also \cite{GY}. 
\bibitem{Sterman} G. Sterman, {\it An Introduction to Quantum Field
Theory}, Cambridge University Press, Cambridge, 1993, pg. 339. 
\bibitem{BP} E. Braaten and R.Pisarski, Nucl. Phys. {\bf B337}, 569
(1987). 
\bibitem{Blaizot} J.-P. Blaizot and E. Iancu, Phys. Rev. Lett. {\bf 76},
3080 (1996). 

\end{thebibliography}
\end{document}